\documentclass[%
 reprint,
superscriptaddress,
 amsmath,amssymb,
 aps,
]{revtex4-2}
\usepackage{braket}
\usepackage{float}
\usepackage{graphicx}% Include figure files
\usepackage{dcolumn}% Align table columns on decimal point
\usepackage{bm}% bold math
\usepackage{hyperref}% add hypertext capabilities

\begin{document}

\title{Sub-Doppler spectra of sodium $D$ lines in a wide range of magnetic field: theoretical study}

\author{Rodolphe MOMIER}
 \email{momier.rodolphe@gmail.com}
\affiliation{Laboratoire Interdisciplinaire Carnot de Bourgogne, UMR CNRS 6303, Université Bourgogne Franche-Comté, 21000 Dijon, France}
\affiliation{Institute for Physical Research, NAS of Armenia, 0203 Ashtarak-2, Republic of Armenia}

\author{Aram V. PAPOYAN}
\affiliation{Institute for Physical Research, NAS of Armenia, 0203 Ashtarak-2, Republic of Armenia}

\author{Claude LEROY}
\affiliation{Laboratoire Interdisciplinaire Carnot de Bourgogne, UMR CNRS 6303, Université Bourgogne Franche-Comté, 21000 Dijon, France}

\date{April 2021}% It is always \today, today,

\begin{abstract}
We compute the interaction of a sodium vapor with a static magnetic field ranging from 0 up to 1 Tesla, which allows to obtain the behavior of all Zeeman transitions as a function of the magnetic field for any polarization of incident laser radiation: $\pi, \sigma^\pm$. We use these results combined with a Fabry-P{\'e}rot microcavity model to describe the transmitted and reflected signal of a sodium vapor confined in a nanometric-thin cell. This allows us to present for the first time high resolution absorption spectra, and provide a complete description of the Zeeman transitions, along with some important peculiarities such as the appearance of guiding transitions and magnetically-induced circular dichroism.  The obtained theoretical results can be used in the upcoming experiments with sodium vapor nanocells. 

\end{abstract}

\maketitle

%\tableofcontents

\section{Introduction}

Alkali metal atoms are recognized as classical objects of research in atomic physics for a number of reasons. In terms of experiment, their convenience is due to the high density of atomic vapor at relatively low temperatures, and the presence of strong optical transitions from the ground state ($D_1$ and $D_2$ lines) in the visible or near-infrared region of the spectrum. From a theoretical point of view, they are attractive due to simplicity of their energy levels caused by the presence of a single electron on the outer shell, which significantly facilitates the calculations, so that spectroscopic data are now very well documented \cite{steck85rb,steck87rb,steckcesium,stecksodium,arimondo,tiecke,gehm}.

Magneto-optical phenomena in alkali metal vapors are of particular interest. Complete understanding of various magneto-optical effects occurring in alkali vapors \cite{budkerREV} is important since they are widely used, for example in fundamental and applied studies of electromagnetically induced transparency \cite{EIT1,EIT2}, Faraday filters \cite{faraday1,faraday2}, optical magnetometry \cite{dijonash3}, laser-frequency locking \cite{locking} and elsewhere. Such phenomena often rely on the peculiarities of the behavior of Zeeman transitions, so that frequency separation of individual transitions is of high importance. Meanwhile for a vapor media, the thermal motion of atoms resulting in Doppler broadening leads to inhomogeneous broadening and overlapping of hyperfine transitions and their Zeeman components, thus limiting its spectroscopic applicability.

Significant sub-Doppler narrowing of atomic transitions can be attained with the use of so-called optical nanocells \cite{SARKISYAN2001201,Sargsyan:17,sargsyanAPL}. Being enclosed in optical cells with a thickness commensurable with the resonant wavelength, alkali metal vapors become a powerful tool for the high-resolution atomic spectroscopy, opening new possibilities for studying magneto-optical processes, where spectral resolution of individual transitions between magnetic sublevels (Zeeman transitions) is essential. Complete determination of the behavior of all the possible individual Zeeman transitions can be done using a well-known theoretical model first provided by Tremblay \textit{et al.} \cite{paper:tremblayPRA}. Importantly, besides frequency splitting in magnetic field, Zeeman transitions undergo also significant probability changes \cite{sargsyanAPL}. This allows to observe the appearance of several peculiarities, such as Guiding Transitions (GT) and two different types of so-called Magnetically-induced Circular Dichroism (MCD) \cite{sargsyanPLA,tonoyanEPL}. Moreover, several efficient techniques have been developed to enhance a spectral visibility of Zeeman transitions while preserving their relative probability scaling, such as derivative of Selective Reflection (dSR) \cite{dijonash2,dijonash3} and Second Derivative of absorption (SD) \cite{arev,jphysbabs,sodiumJETP}. 

Most of the experimental studies was done using rubidium and cesium atomic vapor due to the availability of relatively inexpensive cw diode lasers operating in their resonant wavelength range. 
For these atoms, also the hyperfine splitting is affordably large (the lowest frequency separation between adjacent upper states is $\approx 30$ MHz for $^{85}$Rb).

The magneto-optical effects on sodium $D_1$ and $D_2$ lines have been much less studied. The reason for this is the lack of convenient inexpensive lasers operating in the region of these resonances (589.6 nm and 589.0 nm, respectively), as well as the relatively low vapor density. Although some information concerning Zeeman transitions of sodium exist in the literature \cite{umfer,hori,epjd}, the available information is far from being complete. Meanwhile, recently interest in magneto-optical processes in sodium vapor has increased, in particular, due to the topical problem of mesospheric sodium \cite{Fan:18}.

In this paper, we provide for the first time a comprehensive set of theoretical sub-Doppler spectra of all Zeeman transitions for both lines of sodium in a magnetic field, as well as a thorough description of most of the pecularities that occur. The results we present throughout this paper are qualitatively analogous to the ones obtained for the $D_1$ and $D_2$ line of $^{87}\text{Rb}$ and $^{39}\text{K}$ representing similar quantum systems, therefore we expect total agreement between them and upcoming experiments with Na nanocells.

\section{Theoretical model}
\subsection{Interaction of an alkali vapor with a transverse magnetic field}
The Hamiltonian matrix of alkali atoms interacting with a static magnetic field is expressed as 
\begin{equation}
    \mathcal{H} = \mathcal{H}_0 + \mathcal{H}_m
\end{equation}
where $\mathcal{H}_0$ is the unperturbed (zero-field) Hamiltonian and $\mathcal{H}_m$ is the magnetic contribution given by
\begin{equation}
    \mathcal{H}_m = -\frac{\mu_B}{\hbar}(g_L L_z + g_S S_z + g_I I_z)\, .\label{eq:mag_contribution}
\end{equation}
In relation \eqref{eq:mag_contribution}, $\mu_B$ is the Bohr magneton, $g_L$, $g_S$ and $g_I$ are respectively the orbital, spin and nuclear Landé factors. $L_z$, $S_z$ and $I_z$ are the projection of the orbital, spin and nuclear angular momenta along $z$ which has been chosen as the quantization axis. 
Due to the influence of the magnetic field, the hyperfine energy levels $F$ split into Zeeman sublevels denoted $\ket{F,m_F}$ in the coupled basis, with $m_F$ the magnetic quantum number varying from $-F$ to $+F$. As it has been developed in \cite{paper:tremblayPRA}, the matrix elements of $\mathcal{H}$ obey the two following relations: 

\begin{equation}
     \bra{F,m_F}\mathcal{H}\ket{F,m_F} = E_0(F) - \mu_B g_F m_F B_z \label{eq:diag}
\end{equation}
\small
\begin{align}
   & \bra{F-1,m_F}\mathcal{H}\ket{F,m_F} = \bra{F,m_F}\mathcal{H}\ket{F-1,m_F} \nonumber \\
    &= -\frac{\mu_B B_z}{2}(g_J - g_I)\left(\frac{[(J+I+1)^2-F^2][F^2-(J-I)^2]}{F}\right)^{1/2}\label{eq:nondiag}\\ &\times \left(\frac{F^2-m_F^2}{F(2F+1)(2F-1)}\right)^{1/2}\nonumber\, .
\end{align}
\normalsize
Equation \eqref{eq:diag} gives the diagonal terms of the Hamiltonian, which depend on the projection $B_z$ of $\vec B$ along the quantization axis, the zero-field energy $E_0(F)$ of the $F$ level and the Landé factor $g_F$ associated to the $\ket{F,m_F}$ sublevel (see for example \cite{OPA} or \cite{klingerthese} for a more detailed description). The non-diagonal terms, given by relation \eqref{eq:nondiag}, are non-zero only if they respect the selection rules $\Delta F = \pm 1$, $\Delta m_F = 0$. As a result, the Hamiltonian $\mathcal{H}$ is a block diagonal matrix, each block corresponding to a value of $m_F$. It is important to note that $\mu_B$ has to be chosen with a negative sign for the model to be consistent, and one can refer to \cite{arimondo} for further details.
By diagonalizing the Hamiltonian matrix for each value of the magnetic field, one obtains the eigenvalues which correspond to the energy of each Zeeman sublevel $\ket{F,m_F}$. In the small magnetic field limit ($B_z \ll B_0 = A_{hf}\mu_B^{-1} \approx 63.3$ mT, where $A_{hf}$ is the magnetic dipole constant), the coupled basis is a so-called "good" basis, the Zeeman sublevels split linearly according to \eqref{eq:diag}. For $B \gg B_0 $, it is best to describe the states in the uncoupled basis $\ket{J,m_J,I,m_I}$ where $m_J$ and $m_I$ are projections of the angular momentum $J$ and nuclear momentum $I$, see for example \cite{interf}. In intermediate field ($B \simeq B_0$) one can obtain the energies of the ground states by using the Breit-Rabi formula, as presented in \cite{BR}. 
All the numerical data needed for the computations can be found in the literature, for example in \cite{stecksodium}. All values used for the physical constants are the ones recommended by NIST \cite{nist}. The necessary values are given in table \ref{tab:datasodium}. 
\begin{table}
\centering
\begin{tabular}{|c|c|}
\hline
$\mu_B$  & $1.399 624 493 61(42)\times 10^{10}$ Hz/T \\ \hline
$A_{hf}$ & $885.813 064 4(5)$ MHz     \\ \hline
$g_S$    & $2.00231930436256(35)$     \\ \hline
$g_L$    & $0.99997613$               \\ \hline
$g_I$    & $-0.0008046108(8)$         \\ \hline
\end{tabular}
\caption{Numerical data to compute the interaction of the atomic vapor with a magnetic field.}
\label{tab:datasodium}
\end{table}
\noindent The new state vectors arising due to the influence of the magnetic field are denoted by $\ket{\Psi(F,m_F)}$. They can be expressed as linear combinations of the set of unperturbed state vectors:

\begin{equation}
    \ket{\Psi(F,m_F)} = \sum_{F'}\alpha_{F,F'}(B)\ket{F',m_F}\label{eq:eigenvectors}\, .
\end{equation}
Relation \eqref{eq:eigenvectors} is valid for both ground (denoted by the index $g$ below) and excited states (denoted by the index $e$ below). $\alpha_{F,F'}(B)$ are the magnetic-field-dependent mixing coefficients reflecting the coupling of the Zeeman energy levels due to the magnetic field. The intensity $A_{eg}$ of a transition between two Zeeman sublevels $\Psi(\ket{F_g,m_{F_g}})$ and $\Psi(\ket{F_e,m_{F_e}})$ is proportional to the squared of the so-called "modified transfer coefficients". These coefficients are given by
\begin{align}
  A_{eg} &\propto   a^2[\ket{\Psi(F_e,m_{F_e})};\Psi(F_g,m_{F_g});q] \nonumber \\ & = \left(\sum_{F_e',F_g'}\alpha_{F_e,F_e'}a(F_e',m_{F_e};F_g',m_{F_g};q)\alpha_{F_g,F_g'}\right)^2 \label{eq:aeg}
\end{align}
where $a(F_e,m_{F_e};F_g,m_{F_g};q)$ are the unperturbed transfer coefficients. They have the following form:
\begin{align}
    &a(F_e,m_{F_e};F_g,m_{F_g};q) = (-1)^{1+I+J_e+F_e+F_g-m_{F_e}}\sqrt{2F_e+1}\nonumber\\
    \times & \sqrt{2F_g+1}\sqrt{2J_e+1}
    \begin{pmatrix}
    F_e & 1 & F_g\\
    -m_{F_e} & q & m_{F_g}
    \end{pmatrix}
    \begin{Bmatrix}
    F_e & 1 & F_g\\
    J_g & I & J_e
    \end{Bmatrix}\, .\label{eq:unperturbed}
\end{align}
In relation \eqref{eq:unperturbed}, parenthesis represent a Wigner 3j-symbol and curly brackets a 6j-symbol. The index $q$ reflects the polarization of the laser radiation, $q=0,\pm 1$ for $\pi, \sigma^\pm$. The model allows to study each invidual Zeeman transition of the $D_1$ and $D_2$ lines of any alkali atom, but it also works for other lines with similar structure, as presented in \cite{jqsrt}.  As a last remark, in this work all transitions will be labelled in the basis $\ket{F,m_F}$ for convenience although these quantum numbers form a "good" basis only for low magnetic fields.

\subsection{Transmitted and reflected signal for a vapor of nanometric thickness}
\begin{figure}
    \centering
    \includegraphics[scale=0.56]{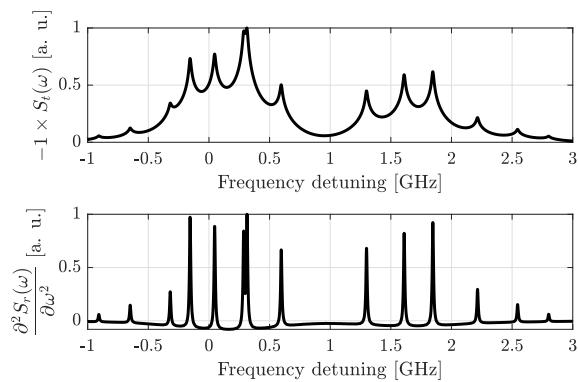}
    \caption{Sodium $D_2$ line spectra for $\pi$-polarized laser radiation. Top: absorption spectrum computed for $L = \lambda/2$. Bottom: dSR spectrum computed for $L = \lambda/2  - 30$ nm.}
    \label{fig:spectre_ex}
\end{figure}

One of the possible ways to obtain sub-Doppler resolution during experiments is to use a nano or micrometric thin-cell filled with an atomic vapor. Such experiments have been thoroughly described in \cite{Sargsyan_2014,Sargsyan_2018,Sargsyan:12,Sargsyan:17,SARKISYAN2001201}. In what follows, we recall briefly the theoretical model that can be used to compute the fields that are reflected $S_r$ and transmitted $S_t$ through a thin cavity (maximum a few wavelengths) seen as a Fabry-Perot interferometer. The reader can refer to \cite{dutierJOSAB} for a complete description of the model. 
The transmitted and reflected signals containing the information on the atomic vapor have the following expressions
\begin{align}
    S_t &\simeq 2t_{wc}t_{cw}^2E_i\mathrm{Re}\left\lbrace I_f - r_wI_b\right\rbrace/|Q|^2 \label{eq:st} \\
    S_r &\simeq2t_{cw}E_i\mathrm{Re}\lbrace r_w[1 - \exp(-2ikL)] \nonumber \\ &\times[I_b - r_w I_f\exp(2ikL)]\rbrace/|Q|^2\, .\label{eq:sr}
\end{align}
where $t$ are transmission coefficients ($w$ stands for windows and $c$ stands for cell), $Q = 1-r_w^2\exp(2ikL)$ is the quality factor of the cavity of length $L$, $r$ is the reflection coefficient (the two windows are here considered identical, and all possible losses due to the windows are neglected). $I_f$ and $I_b$ are called forward and backward integrals of the atomic response
\begin{align}
    I_f &= I_T^{lin} - r_w I_{SR}^{lin}  \\
    I_b &= I_{SR}^{lin} - r_w\exp(2ikL) I_{T}^{lin} \, ,
\end{align}
where $I_T^{lin}$ and $I_{SR}^{lin}$ are the transmitted and reflected signals in the linear regime whose expressions are derived in \cite{dutierJOSAB}. The homogeneous linewidth of the transmitted signal $\gamma = \Gamma_\text{nat}/2$ is a free parameter which can be replaced by $\gamma = \Gamma_\text{nat}/2 + \gamma_f$ to account for additional decoherence processes (for example collisional broadening).
This model, valid for a two-level system having a resonant frequency $\omega_0$, can be extended to an ensemble of two-level systems in order to describe each hyperfine transition having a resonant frequency $\omega_i = \omega_e - \omega_g$ and an amplitude $C_i$ depending mainly on the transition intensity $A_{eg}$, computed using relation \eqref{eq:aeg} (see \cite{klingerthese}). 
 All spectra presented in this paper will be computed for a cell thickness $\lambda/2$ and an absorption linewidth of $100$ MHz. Experimentally, all studies are typically performed around this thickness due to the appearance of Coherent Dicke Narrowing (first observed in the microwave domain \cite{CDN}). Narrowing of the resonance occurs periodically for $L = (2p+1)\lambda/2$ with $p \in \mathbb{N}$. Different lineshapes for the absorption and reflection signal are presented in \cite{dutierJOSAB}. It is relevant to note that for other thicknesses, one obtain assymetric lineshapes arising from different dispersive contributions involved in the expressions of the transmitted and reflected fields. 
As an example of the results provided by this theoretical model, we present on figure \ref{fig:spectre_ex} an absorption and dSR spectrum of $D_2$ line $\pi$-transitions for an external magnetic field $B_z = 50$ mT.

\section{Sodium $D_1$ line}

\begin{figure*}
    \centering
    \includegraphics[scale=0.65]{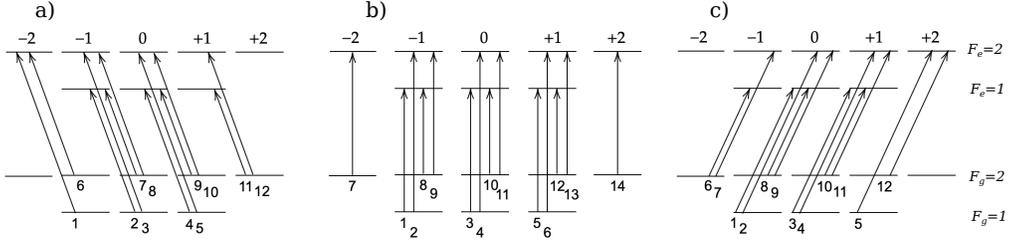}
    \caption{All possible Zeeman transitions for the $D_1$ line of ${}^{23}\text{Na}$ in the basis $\ket{F,m_F}$. a) $\sigma^-$-transitions. b) $\pi$-transitions. c) $\sigma^+$-transitions.}
    \label{fig:schemes}
\end{figure*}

The $D_1$ line of sodium corresponds to the transitions occuring between the states $F_g = 1,2$ of $3^2S_{1/2}$ and $F_e = 1,2$ of $3^2P_{1/2}$. The Zeeman manifold is thus analogous to the one of ${}^{87}\text{Rb}$ or ${}^{39}\text{K}$ $D_1$ line. A $\pi$-polarized incident laser will excite $14$ transitions, whereas left (or right) circularly polarized laser will excite only $12$ transitions. In total for sodium $D_1$ line, $38$ $\ket{F_g,m_{F_g}} \rightarrow \ket{F_e,m_{F_e}}$ Zeeman transitions are possible. A scheme depicting all of them is presented on figure \ref{fig:schemes}.

\noindent Throughout this work, the computations have been performed using $\zeta =  1.771 626 128 8(10)$ GHz the hyperfine splitting between the two ground states $F_g = 1$ and $F_g =2$. In the same manner, $\zeta' = 188.88(26)$~MHz is the hyperfine splitting between $F_e = 1$ and $F_e =2$ \cite{stecksodium}. 

\subsection{Linearly polarized incident radiation}
\begin{figure}
    \centering
    \includegraphics[scale=0.57]{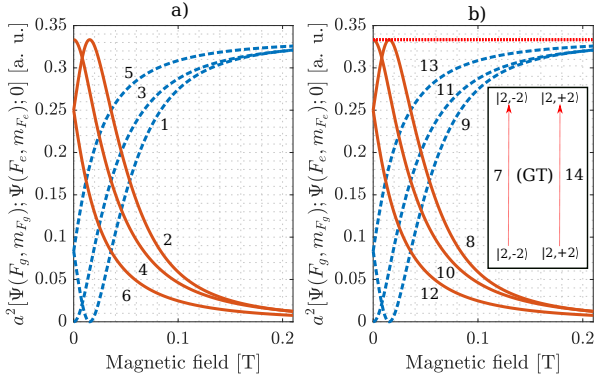}
    \caption{(color online) Sodium $D_1$ line hyperfine transition intensities for $\pi$-polarized laser radiation. a) transitions from $F_g = 1$. b) transitions from $F_g = 2$. The guiding transitions $7$ and $14$ are represented in red (dotted lines). In blue (dashed lines), we represent the transitions whose intensities remain considerable past $B_0 \approx 63.3$ mT, as opposed to the orange curves (continuous lines). Magnetic field varies up to $0.2$ T here for clarity.}
    \label{fig:tr_D1_pi}
\end{figure}

Let us first focus on $\pi$-transitions. As said before, in this case $14$ transitions are possible. All these transitions are so-called "allowed" since they respect the selection rules on $F$: $\Delta F = 0,\pm1$. The transition intensity is proportional to the square of the modified transfer coefficient. The results obtained for $a^2[\ket{\Psi(F_e',m_{F_e})};\Psi(F_g',m_{F_g});+1]$ are presented on figure \ref{fig:tr_D1_pi}. For the sake of clarity to avoid overlapping, we separate the two following cases: transitions labelled $1$ to $6$ with ground state $F_g=1$ are represented on panel a), and transitions labelled $7$ to $14$ with ground state $F_g = 2$ are represented on panel b). 

\begin{figure}
    \centering
    \includegraphics[scale=0.2]{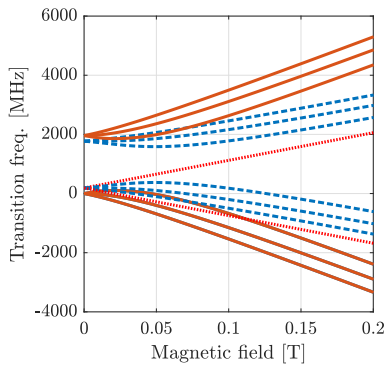}
    \caption{(color online) $\pi$-transition frequencies of sodium $D_1$ line as a function of the magnetic field. The color code corresponds to the one used in figure \ref{fig:tr_D1_pi}. The zero frequency is set to the transition $F_g = 1 \rightarrow F_e = 1$. }
    \label{fig:freq_D1_pi}
\end{figure}

\noindent Several observations can be made from figure \ref{fig:tr_D1_pi}. The amplitudes of transitions $1$, $3$, $5$ (see figure \ref{fig:tr_D1_pi}a),  $7$, $9$, $11$, $13$ and $14$ (see figure \ref{fig:tr_D1_pi}b) remain considerable for $B \gg B_0$, thus $8$ transitions will remain present in the spectrum when the magnetic field is high enough. These transitions will be called r-transitions, where r stands for remaining. Here, they obey the selection $\Delta F = 0$ at low magnetic field ($F$ is not a good quantum number at high magnetic field), and $\Delta m_I = \Delta m_J = 0$ for high magnetic field. The r-transitions $1$, $3$, $5$ occur between $F_g = F_e = 1$, and the others between $F_g = F_e = 2$ . Similarly, transitions $2$, $4$, $6$, $8$, $10$ and $12$ will be called v-transitions, where v stands for vanishing, since their amplitude becomes negligible for $B \gg B_0$.
One can notice that some of the r-transitions show $a^2 \approx 0$ for low magnetic field and experience a huge amplitude growth as the magnetic field increases. For these reason, we call them Magnetically Induced (MI), although they are theoretically allowed by the selection rule $\Delta F = 0,\pm 1$. Another peculiarity is the appearance of "transition cancellations", occuring here at around $15$ mT. This value depends on the different hyperfine splittings. This phenomenon was already mentioned in \cite{paper:tremblayPRA} but no complete description was done. 
Precise discussion of these cancellations is out of the scope of this paper, however we have provided a thorough analysis and a way to determine precisely the magnetic field values leading to dipole moment cancellation in \cite{arturJOSAB,jqsrt,arturarxiv}. Using the relations provided in these works, we can determine that transitions $1$ and $9$ experience dipole moment cancellation for 
\begin{equation}
B = \frac{1}{\mu_B}\frac{3\zeta\zeta'}{(3g_I - 4g_L + g_S)\zeta + 3(g_I - g_S)\zeta'} \approx 15.32~\text{mT}\,.
\end{equation}
The last peculiarity that appears is the presence of two r-transitions ($7$ and $14$) having a constant amplitude throughout the whole range of magnetic field. These transitions occur between uncoupled states (following the definition of relation \eqref{eq:eigenvectors}): such states do not experience any mixing with their neighbors (with the same $m_F$ value), thus leading to magnetic-field-independent transition amplitudes. As shown on the inset of the right panel of figure \ref{fig:tr_D1_pi}b, these transitions are called Guiding Transitions (GT). For further analysis, we present on figure \ref{fig:freq_D1_pi} the frequencies $\omega_{eg}(B_z) = (E_e(B_z)- E_g(B_z))/\hbar$ of all the possible $\pi$-transitions. 

\begin{figure}
    \centering
    \includegraphics[scale=0.5]{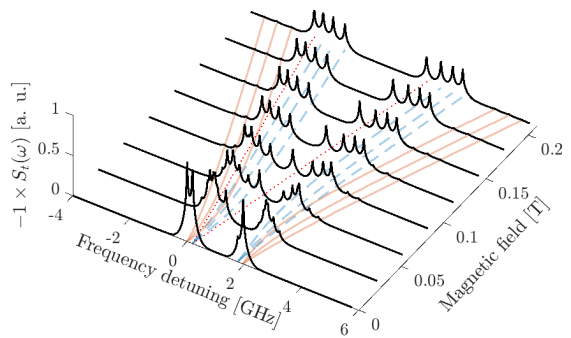}    \caption{(color online) Theoretical absorption spectra of the $D_1$ line of sodium for $\pi$-polarized incident laser radiation. The magnetic field intensity varies from $0$ to $0.21$ T with a step of $30$ mT. The shadow lines on the $(x,y)$ plane are the same transition frequencies shown on figure \ref{fig:freq_D1_pi}. Each component has a Full Width at Half Maximum (FWHM) of $\omega_{\text{FWHM}} \approx 100$ MHz. The zero frequency is set to the transition $F_g = 1 \rightarrow F_e = 1$.}
    \label{fig:abs_D1_pi}
\end{figure}

\noindent From figure \ref{fig:freq_D1_pi}, one can notice the following things: the two guiding transitions (represented in red dotted lines) occur between uncoupled states having for only difference their $m_F$ values ($m_{F_g} = m_{F_e} = \pm 2$), their frequency shifts are perfectly linear with respect to $B_z$ and only differ by sign. Going back to the model of section 1, the uncoupled states give rise to $1\times 1$ blocks in $\mathcal{H}$ and thus their Zeeman splitting is perfectly described by relation \eqref{eq:diag}, hence it is quite straightforward to show that their slope is 
\begin{equation}
s_{\pm,GT}^{D_1^{(\pi)}} \approx \pm\frac{2\mu_B}{3}\, ,\label{eq:slopeGTD1}
\end{equation}
under the approximation $g_S \approx 2$, $g_L \approx 1$ and $g_I \ll g_J$. Numerically, by using the values provided in table \ref{tab:datasodium}, we obtain $s_{\pm,GT}^{D_1^{(\pi)}} = \pm 0.935269$ MHz/G. Each group of r-transitions is thus driven by one of the GTs, more precisely r-transitions $1$, $3$ and $5$ are driven by the GT labelled $7$ having a slope $s_{-,GT}^{D_1^{(\pi)}} = -2\mu_B/3$, and r-transitions $9$, $11$ and $13$ are driven by the GT labelled $14$ having a slope $s_{+,GT}^{D_1^{(\pi)}} = 2\mu_B/3$, with $\mu_B$ chosen to be negative as mentioned earlier. Driven means here that each individual transition belonging to a given group is bounded in amplitude by the one the Guiding Transition, and all the frequency shifts asymptotically tend to a linear behavior when $B \gg B_0 $ is high enough, with the slope being $ s_{\pm,r}^{D_1^{(\pi)}}= s_{\pm,GT}^{D_1^{(\pi)}}$, as written in relation \eqref{eq:slopeGTD1}. Due to the absence of coupling between states, the modified transfer coefficient of a GT occuring between two uncoupled states $\ket{F_g,m_{F_g}} \rightarrow \ket{F_e,m_{F_e}}$ is exactly equal to the unperturbed transfer coefficient $a^2(F_e,m_{F_e};F_g,m_{F_g},q)$. For transitions $7$ and $14$, we obtain 
\begin{equation}
a(2,\pm 2;2,\pm2;q=0) = \pm \frac{1}{\sqrt{3}} \rightarrow a^2 = \frac{1}{3}\label{eq:unmod} \, .
\end{equation}
Past $B_0$, the 8 r-transitions remain in the spectrum, which is a manifestation of the hyperfine Paschen-Back regime. This behavior has been observed experimentally for cesium, rubidium and potassium (for example in \cite{optcommarmen}). As mentioned before, experimental results are harder to provide due to the lack of diode lasers operating in the wavelength range of sodium. However, a sodium nanocell has been developed and is available in our laboratory \cite{sodiumJETP}.  A set of theoretical absorption spectra for the all $\pi$-transitions of sodium $D_1$ line is presented on figure \ref{fig:abs_D1_pi} for visualization. 

\begin{figure}
    \centering
    \includegraphics[scale=0.5]{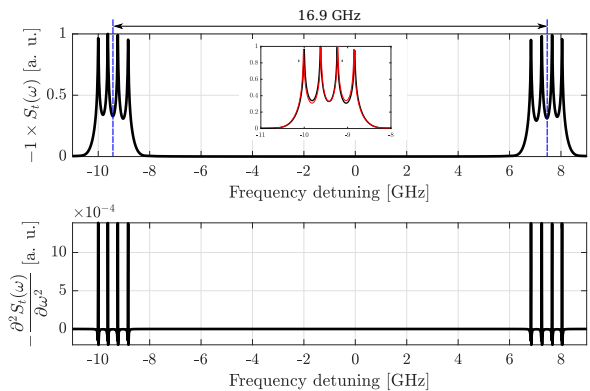}
    \caption{(color online) Theoretical absorption spectra of the $D_1$ line of sodium for $\pi$-polarized incident laser radiation in the hyperfine Paschen-Back regime ($B=0.9$ T). Top: transmitted signal. Bottom: second derivative (SD). Inset: "mirror" overlap of the two groups when the high-frequency group is computed as a function of $-\omega$. The top curve has been normalized so that the maximum amplitude is $1$.}
    \label{fig:abs_D1_pi_HPB}
\end{figure}

\noindent All the previous statements are visible on figure \ref{fig:abs_D1_pi}. The r-transitions can be separated in two groups all tending to the same amplitude and a complete symmetry between the red and blue wings of the spectrum appears when $B$ is high enough. Although becoming invisible, the v-transition frequency shifts also exhibit a linear behavior at high magnetic fields, tending to a slope twice bigger than for the r-transitions. Precisely, $s_{\pm,v}^{D_1^{(\pi)}} = 2s_{\pm,r}^{D_1^{(\pi)}} \approx \pm 4\mu_B/3$ using the same approximations. It is usually estimated that complete Hyperfine Paschen-Back regime is reached when a magnetic field $B > 10B_0$ is applied. This correspond here to a magnetic field $B_z \approx 0.63$ T. As a comparison, for rubidium 87, $B_0 \approx 0.24$ T  \cite{steck87rb} and for cesium, $B_0 \approx 0.16$ T \cite{steckcesium}. Theoretically, this makes sodium, along with potassium 39 \cite{klingerthese,tiecke} more convenient than cesium or rubidium for the experimental study of magnetic-optical processes occuring in the complete hyperfine Paschen-Back since it is reached for a much weaker value of external magnetic field. However, since the natural linewidth of sodium $D_1$ line $\Gamma_\text{nat}/2\pi = 9.765$ MHz \cite{stecksodium} is nearly twice bigger than for the $D_1$ lines of cesium or rubidium, one expects to obtain significantly broader lines when studying sodium. The linewidth is also affected by the inhomogeneous Doppler broadening $\Gamma_D$ given by
\begin{equation}
\Gamma_D = \omega_0 \sqrt{\frac{8k_B T \ln 2 }{mc^2}}\, ,\label{dopplerbroadening}
\end{equation} 
(see \cite{AMP}) bigger for the $D_1$ line of sodium than in the case of rubidium or cesium.
In relation \eqref{dopplerbroadening}, $\omega_0$ is the central angular frequency of the transition, $m$ is the atomic mass,  $T$ is the temperature of vapor, $c$ is the speed of light and $k_B$ is the Boltzmann constant. However, performing experiments with nanocells allows to reach sub-Doppler spectral resolution  \cite{SARKISYAN2001201}. On figure \ref{fig:abs_D1_pi_HPB}, we present a theoretical absorption spectrum of sodium $D_1$ $\pi$-transitions, computed with the same numerical parameters as before, as long with the second derivative spectrum (SD). 

\noindent Due to the linewidth $\omega_\text{FWHM}\approx 100$ MHz, even for a magnetic field $B_z \gg 10 B_0$, the peaks are not totally resolved hence resulting in a variation of the peaks' amplitude.  Experimentally, as demonstrated on the bottom panel of figure \ref{fig:abs_D1_pi_HPB}, SD would allow to obtain completely resolved peaks, all having the same amplitude. DSR would also be suitable here. These techniques allow to obtain much narrower resonances (two to three times, see \cite{klingerthese}). One can also see that all the peaks are evenly spaced, which is a manifestation of the linear behavior of the frequency shifts mentioned above and a good sign HPB regime is reached. Moreover, as shown by the inset, the red and blue wing of the spectrum are completely symmetric. The frequency detuning $\Delta\omega_r^{D_1^{(\pi)}}$ between the two groups of $4$ r-transitions (their center of gravity, see the dashed blue lines on figure \ref{fig:abs_D1_pi_HPB}) for $B_z \gg B_0$ can be estimated roughly by performing the difference between the two GTs, thus 

\begin{equation}
\Delta\omega_r^{D_1^{(\pi)}}\approx \left| \frac{4\mu_BB_z}{3}\right|\, .
\end{equation}
For a magnetic field $B_z = 0.9$ T, we obtain $\Delta\omega_r^{D_1^{(\pi)}} \approx 16.8$ GHz which is in excellent agreement with the value $16.9$ GHz computed numerically and shown on figure \ref{fig:abs_D1_pi_HPB}.

\subsection{Circularly polarized incident radiation}
\begin{figure}
    \centering
    \includegraphics[scale=0.57]{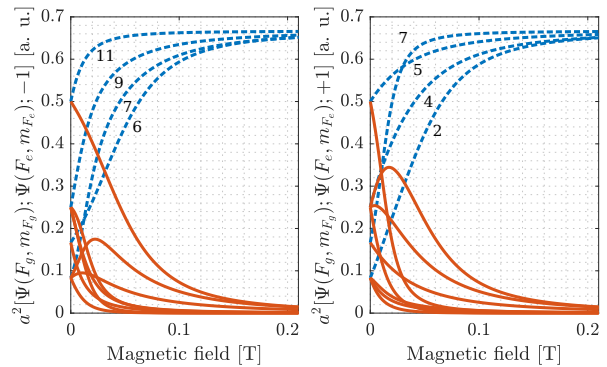}
    \caption{(color online) Sodium $D_1$ Zeeman transition intensities for circularly polarized laser radiation. a) $\sigma^-$-transitions. b) $\sigma^+$-transitions.  Labelling is provided only for the non-vanishing transitions (dashed blue lines) and is the same as on panels a) and c) of figure \ref{fig:schemes}.}
    \label{fig:tr_D1_spsm}
\end{figure}

All the possible $\sigma^-$ (resp. $\sigma^+$) transitions of the $D_1$ line of sodium are schematized on panel a) (resp. panel b)) of figure \ref{fig:schemes}, and their associated intensities are presented on figure \ref{fig:tr_D1_spsm}. 

\begin{figure}
    \centering
    \includegraphics[scale=0.5]{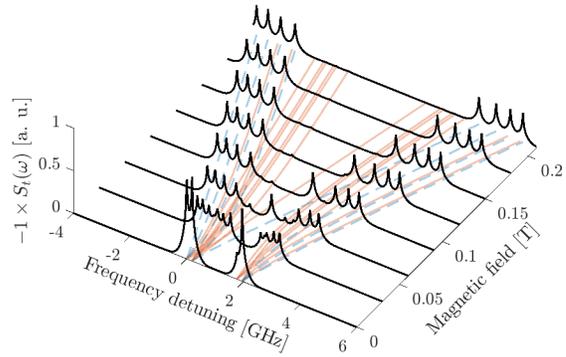}
    \caption{(color online) Theoretical absorption spectra of sodium $D_1$ line for a magnetic field varying from $0$ to $0.21$ T with a step of $0.03$ T in case of simultaneous $\sigma^+$ and $\sigma^-$ excitation. Blue and orange shadow lines in the $(x,y)$ plane represent the transition frequencies. The zero detuning is set to the transition $F_g = 1 \rightarrow F_e = 1$. As before, each peak has a FWHM of approximately $100$ MHz.}
    \label{fig:abs_D1_sigma}
\end{figure}

\noindent As a first remark, no $\sigma^\pm$ has a constant amplitude throughout the whole magnetic field range, which is due to the fact that no transition occurs between uncoupled states. At best, only the excited state is uncoupled (transitions $1$ and $6$ for $\sigma^-$, and their "symmetric" transitions $5$ and $12$ for $\sigma^+$), but this is not enough to avoid the magnetic-field dependence of the transition intensities, thus there is no guiding phenomenon as it happens for $\pi$-transitions. Although being presented in arbitrary units, the amplitude scale used for figure \ref{fig:tr_D1_spsm} is the same as for the $\pi$-transitions, allowing to compare the three cases. Each circular polarization gives rise to a group of four r-transitions, all tending to an amplitude ($\approx 0.66$ a. u.) twice bigger than the amplitude of the $\pi$-GTs ($\approx 0.33$ a. u.). The r-transitions obey the selection rules $\Delta m_I = 0$, $\Delta m_J = \pm 1$ depending on the sign of the incident polarization. A set of absorption spectra is presented on figure \ref{fig:abs_D1_sigma}.

\begin{figure}
    \centering
    \includegraphics[scale=0.52]{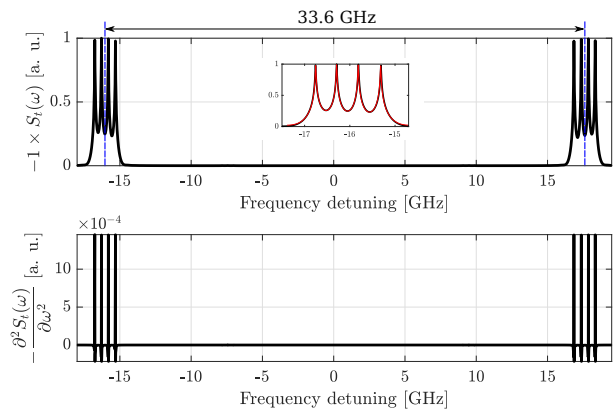}
    \caption{(color online) Theoretical absorption spectrum of the $D_1$ line of sodium for simultaneous $\sigma\pm$ polarized incident laser radiation in the hyperfine Paschen-Back regime ($B=0.9$ T). Top: transmitted signal. Bottom: second derivative (SD). Inset: "mirror" overlap of the two groups when the high-frequency group is computed as a function of $-\omega$. The top curve has been normalized so that the maximum amplitude is $1$.}
    \label{fig:abs_D1_spsm_HPB}
\end{figure}

\noindent Due to the absence of GTs for circularly polarized incident radiation, less pecularities occur in this case. One can also note the absence of dipole moment cancellation, as was observed in \cite{arturJOSAB,jqsrt}. However, some observations can still be made from figure \ref{fig:abs_D1_sigma}. Here, the behavior of the r-transitions is a bit different than before: for $\sigma^+$ polarization, the frequency shifts all tend to a linear behavior of slope twice bigger than for the $\pi$-r-transitions. For example, we take a look at transition $11$ ($\sigma^-$) occuring between the ground state $\ket{2,2}$ and the excited state $\ket{1,1}$ Using relation \eqref{eq:diag}, we can easily determine that 
\begin{equation}
E_g = \zeta - \mu_B B_z\, .\label{eq:eg}
\end{equation}
For the excited state, we only compute the block corresponding to $m_{F_e} = 1$ using relations \eqref{eq:diag} and \eqref{eq:nondiag}. This leads to 
\begin{equation}
H_e = \begin{pmatrix}
X & -\sqrt{3}X \\ 
-\sqrt{3}X & \zeta'-X
\end{pmatrix} 
\end{equation}
where we denoted $X = \mu_B B_z / 6$. Diagonalizing $H_e$ and performing an asymptotic expansion allows to obtain the energy $E_e$ of state $\ket{1,1}$ for $B_z \gg B_0$:
\begin{align}
E_e &= \frac{\zeta' + [\zeta'^2 - 4\zeta'X + 16X^2]^{1/2}}{2} \nonumber \\ &\underset{X\rightarrow+\infty}{\sim} 2X + \frac{\zeta'}{4} + \frac{3\zeta'^2}{64X} + \mathcal{O}\left(\frac{1}{X^2}\right) \, .\label{eq:ee}
\end{align}
The difference between \eqref{eq:ee} and \eqref{eq:eg} leads to $8X + \zeta'/4 - \zeta + \mathcal{O}{(1/X)}$, thus we obtain the asymptotic slope of the frequency shift of transition $11$ ($\sigma^-$) and by extension of all $\sigma^\pm$ r-transitions:
\begin{equation}
s_{\pm,r}^{D_1^{(\sigma^\pm)}} \approx \mp \frac{4\mu_B}{3}\, .\label{eq:slopesigmad1}
\end{equation}
This method was used just before for the $D_1$ $\pi$-transitions and will be used throughout all this paper. As we can see on figure \ref{fig:abs_D1_sigma}, it can be demonstrated that the v-transitions (orange continuous lines) that are not overlapped with any r-transition (blue dashed lines) have a frequency shift of slope twice smaller than the value presented in relation \eqref{eq:slopesigmad1}. We present two spectra for $B\gg B_0$ on figure \ref{fig:abs_D1_spsm_HPB}. 

\noindent The numerical parameters used for the computation of the spectra are the same as the ones used before in the case of $\pi$-transitions. On the top panel, the peaks are overlapped due to the linewidth. However, clear manifestation of the hyperfine Paschen-Back regime is visible on the bottom panel where all the peaks are evenly spaced and have the same amplitude. Due to the fact that the slope of the frequency shifts corresponding to the two groups of r-transitions is twice bigger, the frequency detuning between them can also be roughly estimated by
 
\begin{equation}
\Delta\omega_r^{D_1^{(\sigma)}} \approx \left| \frac{8\mu_BB_z}{3}\right|\, .
\end{equation}
Here, we obtain $\Delta\omega_r^{D_1^{(\sigma)}}\approx  33.59$ GHz which is in perfect agreement with the value $33.6$ GHz measured on figure \ref{fig:abs_D1_spsm_HPB}. In this section, we have provided a complete description of the influence of the magnetic field on the behavior of the Zeeman transitions of sodium $D_1$ line for the three main types of incident laser radiation. We will now analyze deeply what happens for the $D_2$ line, where much more transitions and different phenomena arise.

\section{Sodium $D_2$ line}
The $D_2$ line of sodium corresponds to the transitions occuring between the states $3^2S_{1/2}$ and $3^2P_{3/2}$. Due to the bigger value of $J_e$, much more transitions are possible than for the $D_1$ line. In total, $68$ transitions are possible. Since natural sodium is only composed of one isotope of nuclear spin $I = 3/2$, the hyperfine manifold is still simpler than the one of natural rubidium (two isotopes: ${}^{85}\text{Rb}$ ($I = 5/2$) and ${}^{87}\text{Rb}$ ($I = 3/2$)) or cesium (only one isotope but $I=7/2$), thus leading to spectra which are simpler to analyze. For the numerical computations, we considered the hyperfine splittings $\alpha' = 15.809(80)$ MHz between $F_e = 0$ and $F_e  =1$, $\beta' = 34.344(69)$ MHz between $F_e = 1$ and $F_e = 2$, as well as $\delta' = 58.326(43)$ MHz between $F_e = 2$ and $F_e = 3$.

\subsection{Linearly polarized incident radiation}

\begin{figure*}
    \centering
    \includegraphics[scale=0.45]{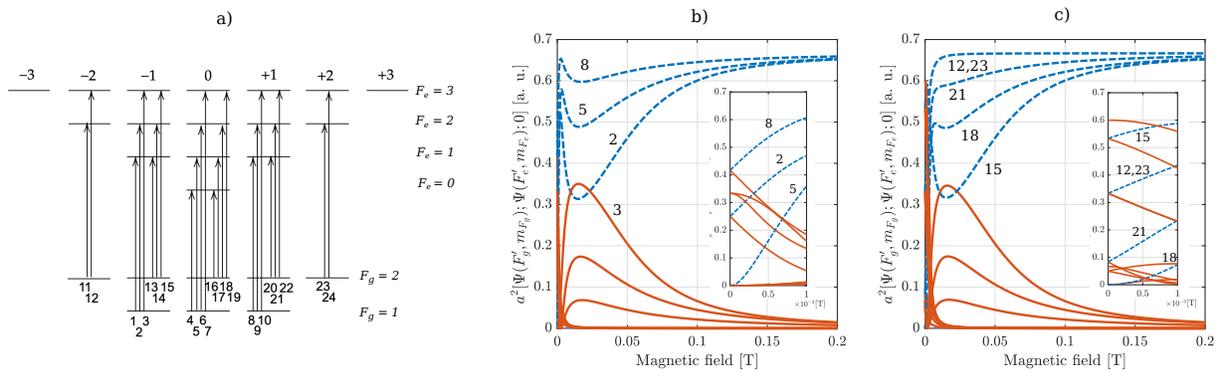}
    \caption{(color online) a) scheme of all possible $\pi$-transitions for the $D_2$ line of sodium. b) transitions from $F_g = 1$. c) transitions from $F_g = 2$. Respective behaviors for very low magnetic field are shown in the insets. The v-transitions are represented in orange (continuous lines) and the r-transitions are represented in blue (dashed lines).}
    \label{fig:tr_D2_pi}
\end{figure*}

$24$ $\pi$-transitions are possible for the $D_2$ line of sodium. A complete scheme as well as the transition intensities are presented on figure \ref{fig:tr_D2_pi}. In this case, $F_g = 1,2$ and $F_e = 0,1,2,3$ \cite{stecksodium}.

\begin{figure}
    \centering
    \includegraphics[scale=0.5]{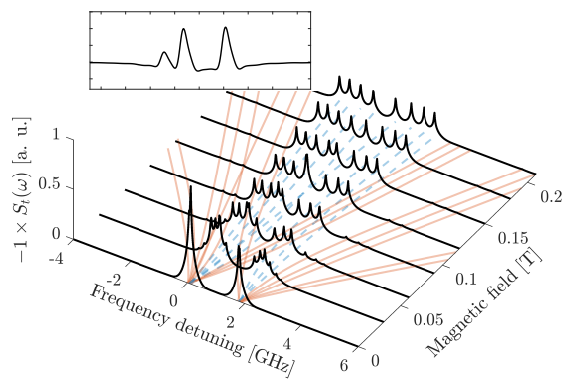}
    \caption{(color online) Theoretical absorption spectra of the $D_2$ line of sodium for $\pi$-polarized incident laser radiation. The magnetic field intensity varies from $0$ to $0.21$ T with a step of $30$ mT. The zero frequency is set to the transition $F_g = 1 \rightarrow F_e = 0$. The inset shows the SD of the three peaks $F_g = 1 \rightarrow F_e = 0,1,2$.}
    \label{fig:abs_D2_pi}
\end{figure}

\noindent The behavior of the $\pi$-transitions of sodium $D_2$ line is quite different from the $D_1$ line. One quickly notices the absence of transitions between uncoupled states thus the absence of guiding transitions. This is due to absence of states such that $F_g = 3$. Overall, the general shape is different, and this is due to the fact that the states experience "more coupling" (for a given spectroscopic state in the case of $D_1$ line, the magnetic field can couple at most levels two $F$-levels, whereas for the $D_2$ line the coupling goes up to three $F$ levels. Many transition intensities are very small throughout the whole range of magnetic field and most of them quickly vanish, as we can notice on the insets of figure \ref{fig:tr_D2_pi}.
Using the same denomination, we call r-transitions the transitions $2$, $5$, $8$, $12$, $15$, $18$, $21$ and $23$ since they obey the selection rules $\Delta m_I = \Delta m_J = 0$, and all the others will be denoted v-transitions. It is worth noticing that for the $D_1$ line, we could exhibit the fact that all r-transitions obeyed $\Delta F = 0$. No such rule in the uncoupled basis can be exhibited here. 
In the previous part, we called MI the transitions whose intensity was $0$ for a very small magnetic field. On figure \ref{fig:tr_D2_pi}, this phenomenon is clearly visible for r-transitions $5$ and $18$. For the sake of clarity, labelling is not presented here for the v-transitions although it is completely possible to associate each curve to a given transition.
However, this notion of Magnetically Induced transitions can be completed. Here, we will also denote transitions $3$, $7$, $10$ and $16$ as Magnetically Induced. All of them are v-transitions and their amplitude tends to $0$ for $B_z \rightarrow 0$, except for transition $3$ ($\ket{1,-1}\rightarrow\ket{3',-1}$) which experiences a huge increase in amplitude. They are called magnetically induced due to the coupling of states by $B_z$ making so-called "forbidden" transitions ($\Delta F = \pm 2$) possible. 
Here, we have $3$ r-transitions arising from $F_g = 1$ and $5$ from $F_g = 2$ (transitions $12$ and $23$ are exactly overlapped on figure \ref{fig:tr_D2_pi}). The total number of $\pi$-r-transitions is the same as for the $D_1$ line meaning there are much more v-transitions here (16 compared to 6). On figure \ref{fig:abs_D2_pi}, we present a set of absorption spectra along with the transition frequencies to observe their behavior as the magnetic field increases.

\noindent Since the hyperfine splittings of the states $3^2P_{3/2}$ are much smaller for sodium than for rubidium or cesium \cite{stecksodium,steck87rb,steck85rb,steckcesium}, the peaks are in general much closer here. At zero-field the 6 peaks remain completely overlapped. Using SD or dSR is a good way to recover the spectral information, as demonstrated on the inset of figure \ref{fig:abs_D2_pi} where the relative transition intensities are in perfect agreement with the oscillator strengths presented in \cite{stecksodium} ($S_{10} = 1/6$, $S_{11} = S_{12} = 5/12$). 

\begin{figure}
    \centering
    \includegraphics[scale=0.52]{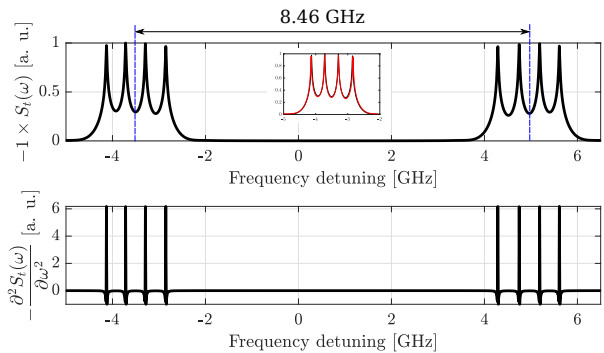}
    \caption{(color online) Theoretical absorption spectrum of the $D_2$ line of sodium for $\pi$-polarized excitation in the hyperfine Paschen-Back regime ($B=0.9$ T). Top: transmitted signal. Bottom: second derivative (SD). Inset: "mirror" overlap of the two groups when the high-frequency group is computed as a function of $-\omega$. The top curve has been normalized so that the maximum amplitude is $1$.}
    \label{fig:abs_D2_pi_HPB}
\end{figure}

At first, for the same values of magnetic field as before, one notices that all the transitions experiencing the biggest frequency shift with respect to the magnetic field are the 16 v-transitions. They can be divided into three groups of same frequency shift (in absolute value). The frequency shifts of the two groups of r-transitions tend asymptotically to a linear behavior of slope
\begin{equation}
s_{\pm,r}^{D_2^{(\pi)}} \approx \pm \frac{\mu_B}{3} = \frac{s_{\pm,r}^{D_1^{(\pi)}}}{2}\,.\label{rslopeD2}
\end{equation}
All the slopes are derived using similar procedures used in relations \eqref{eq:eg} to \eqref{eq:slopesigmad1}. 
The v-transitions undergo much bigger frequency shifts and can be divided into three groups having asymptotically the same slope in absolute value. Under the approximations $g_S \approx 2$, $g_L \approx 1$ and $g_I \ll g_J$, the first group of v-transitions (in terms of proximity with the r-transitions) experiences a frequency shift of asymptotic slope $ \pm \mu_B $. The frequency shifts of the second group of v-transitions have an asymptotic slope $\pm 5\mu_B/3$. This slope for the last group is $\pm 3\mu_B$. This is coherent with figure \ref{fig:abs_D2_pi} where we can see the last group of v-transitions experience the biggest shifts (more than $4$ MHz/G, 9 times more than the r-transitions).
Numerically, we obtain respectively $\pm 1.40061$, $\pm 2.29035$ and $\pm4.15417$ MHz/G. A spectrum for $B_z \gg B_0$ is presented on figure \ref{fig:abs_D2_pi_HPB}.

\noindent As before, the bottom panel shows that SD is convenient to obtain complete resolution of peaks. Hyperfine Paschen-Back is reached since all the peaks are evenly spaced and tend to the same amplitude. It is again possible to estimate the frequency detuning between the two groups of r-transitions as

\begin{equation}
\Delta\omega_r^{D_2^{(\pi)}} \approx \left| \frac{2\mu_BB_z}{3}\right|\, .
\end{equation}
Here, we obtain $\Delta\omega_r^{D_2^{(\pi)}} \approx 8.40$ GHz (by numerical computation, we obtain $8.46$ GHz. The small variation is due to the fact that among all approximations, the influence of the hyperfine splittings is neglected. 

\subsection{Circularly polarized incident laser radiation}
\begin{figure*}
    \centering
    \includegraphics[scale=0.6]{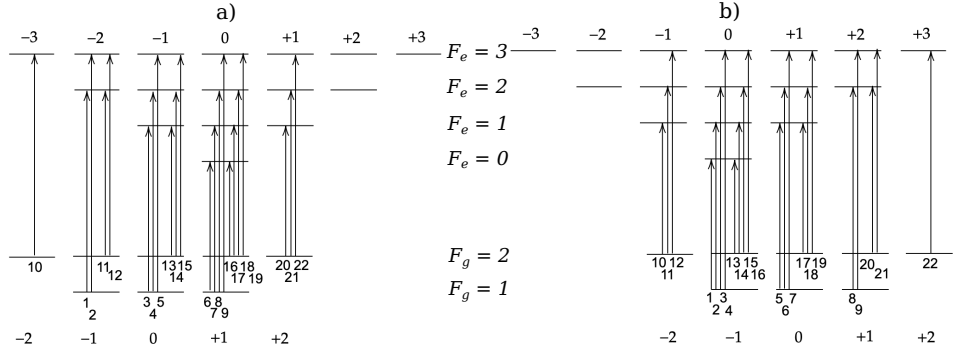}
    \caption{All possible $\sigma^\pm$ Zeeman transitions for the $D_2$ line of sodium in the basis $\ket{F,m_F}$. a) $\sigma^-$-transitions. b) $\sigma^+$-transitions.}
    \label{fig:schemes_D2_sigma}
\end{figure*}

$44$ Zeeman transitions are possible in case of circularly polarized radiation for the $D_2$ line of sodium. All these possible $\sigma^\pm$-transitions are schematized on figure \ref{fig:schemes_D2_sigma}. 

\begin{figure}
    \centering
    \includegraphics[scale=0.57]{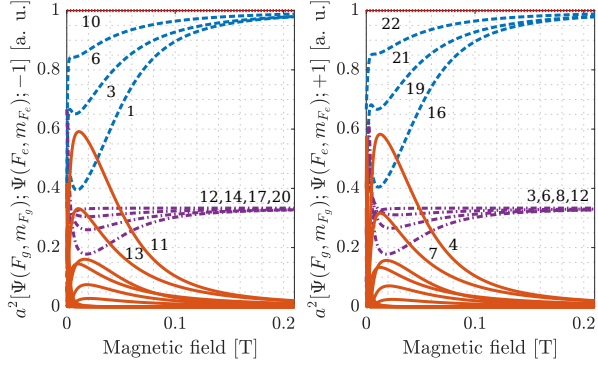}
    \caption{(color online) Sodium hyperfine transition intensities for circularly polarized laser radiation. a) $\sigma^-$-transitions. b) $\sigma^+$-transitions. Past $B_0 \approx 0.063$ T, two different groups of transitions have an amplitude that remains considerable. These two different groups are represented in blue (dashed lines) and purple (dash-dotted), and the vanishing transitions are represented in orange. Labelling is consistent with figure \ref{fig:schemes_D2_sigma}.}
    \label{fig:tr_D2_spsm}
\end{figure}

\noindent One directly sees on the manifold presented on figure \ref{fig:schemes_D2_sigma} that, due to the selection rule $\Delta m_F = \pm 1$, transitions are again possible between uncoupled states. These transitions are labelled $10$ ($\sigma^-$) and $22$ ($\sigma^+$). They occur between states 
$\ket{2,\pm 2}$ and $\ket{3,\pm 3}$ depending on the polarization and obey $\Delta F = 1$. They are again called guiding transitions and their constant amplitude for any magnetic field is clearly visible on figure \ref{fig:tr_D2_spsm}.

\noindent For each different polarization appears a group of r-transitions ($1$, $3$, $6$ and $10$ for $\sigma^-$ and $16$, $19$, $21$, $22$ for $\sigma^+$) denoted r1-transitions.
Transition $10$ ($\sigma^-$) acts a guiding transition meaning it is an amplitude bound for all its associated r1-transitions. For $\sigma^+$, $16$, $19$ and $21$ are all bounded in amplitude by transition $22$.
As we did in relation \eqref{eq:unmod}, we can prove that for both of these transitions, the squared modified transfer coefficients are always equal to 1. Two other groups of r-transitions, denoted r2-transitions, represented in purple (dash-dotted lines), appear and all tend to an amplitude equal to a third the amplitude of the r1-transitions. For both polarization, r2-transitions tend to reach their maximum amplitude much faster than r1-transitions. In orange are represented the vanishing transitions (v-transitions) for each polarizations. Apart from transitions $11$ and $13$ ($\sigma^-$) and transitions $4$ and $7$ ($\sigma^+$), all the v-transitions have overall smaller amplitudes than all r1 and r2-transitions. 

\begin{figure}
    \centering
    \includegraphics[scale=0.5]{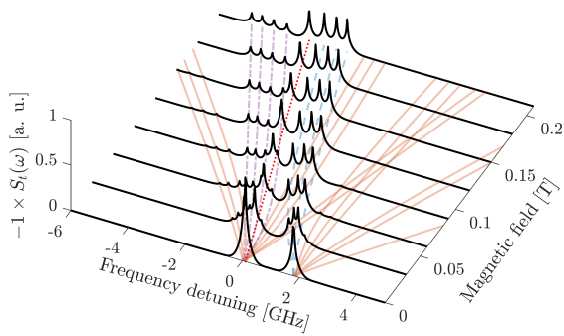}
    \caption{(color online) Theoretical absorption spectra of the $D_2$ line of sodium for $\sigma^-$-polarized incident laser radiation. The magnetic field intensity varies from $0$ to $0.21$ T with a step of $30$ mT. The zero frequency is set to the transition $F_g = 1 \rightarrow F_e = 0$.}
    \label{fig:abs_D2_sm}
\end{figure}

\begin{figure}[h]
    \centering
    \includegraphics[scale=0.5]{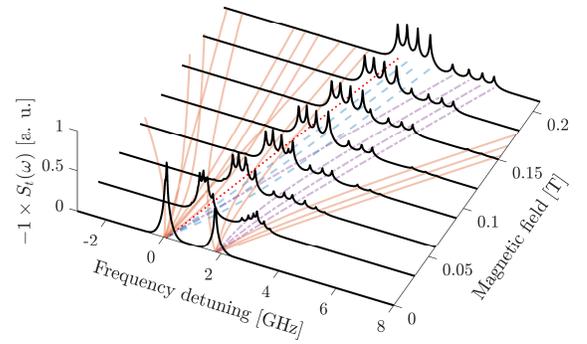}
    \caption{(color online) Theoretical absorption spectra of the $D_2$ line of sodium for $\sigma^+$-polarized incident laser radiation. The zero frequency is set to the transition $F_g = 1 \rightarrow F_e = 0$.}
    \label{fig:abs_D2_sp}
\end{figure}

\noindent The $22$ possible $\sigma^-$-transitions are visible on figure \ref{fig:abs_D2_sm}. The GT (labelled $10$ on figure \ref{fig:schemes_D2_sigma}a) has a constant amplitude and the three other r1-transitions are also driven by the GT in terms of frequency shift. The guiding transition has a perfectly linear frequency shift of slope 

\begin{equation}
s_{GT}^{D_2^{(\sigma^-)}} \approx \mu_B ,\label{eq:slopeGTD2sp}
\end{equation}
under the usual approximations, the numerical value being $\approx -1.39958$ MHz/G while $\mu_B = -1.39962$ MHz/G.
The r2-transitions, having an amplitude three times smaller, experience a frequency shift of approximately $5\mu_B/3$ for $B_z$ high enough. Numerically, we obtain approximately $-2.331$ MHz/G. As for the previous cases, the v-transitions experience much bigger frequency shifts, reaching again as much as approximately $3\mu_B = -4.199$ MHz/G (numerically, $-4.202$ MHz/G). The behavior of the $\sigma^+$ is reversed compared to $\sigma^-$, as we can see on figure \ref{fig:abs_D2_sp}.

\begin{figure}[h]
    \centering
    \includegraphics[scale=0.5]{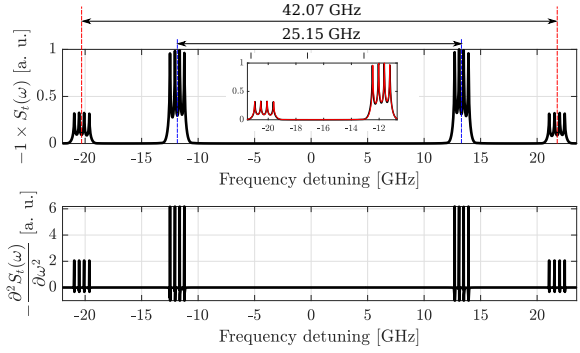}
    \caption{(color online) Theoretical absorption spectrum of the $D_2$ line of sodium for simultaneous $\sigma^\pm$ polarized excitation in the hyperfine Paschen-Back regime ($B=0.9$ T). Top: transmitted signal. Bottom: second derivative (SD).}
    \label{fig:abs_D2_sigma_HPB}
\end{figure}

\noindent The guiding transition ($22$ in this case) experiences a frequency shift $s_{GT}^{D_2^{(\sigma^+)}} = -s_{GT}^{D_2^{(\sigma^-)}}$ and so do all the r1-transitions when $B$ is high enough so that the frequency shift becomes linear. The r2-transitions will also analogously experience a linear frequency shift $-5\mu_B/3$ for $B_z$ (numerically, $\approx 2.331$ MHz/G). When the hyperfine Paschen-Back regime is reached ($B_z > 10B_0$), only the r1 and r2-transitions of each polarization remain visible in the spectrum ($16$ peaks in total). This is clearly demonstrated on figure \ref{fig:abs_D2_sigma_HPB} where the spectrum is presented in case of simultaneous $\sigma^\pm$ excitation for $B_z = 0.9$ T.

\begin{figure}[h]
    \centering
    \includegraphics[scale=0.58]{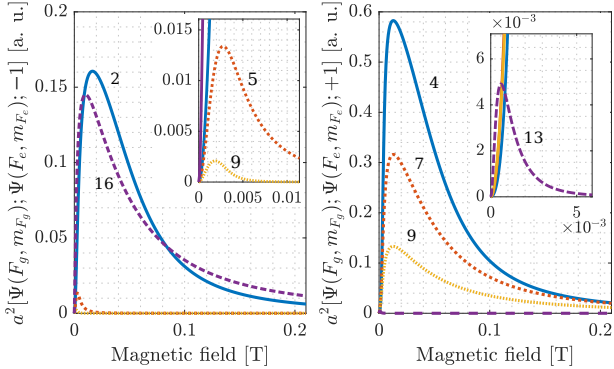}
    \caption{(color online) Transition intensities obeying $\Delta F = \pm 2$. a) $\sigma^-$-transitions. b) $\sigma^+$-transitions. The insets show the very-low-field behavior of transitions having a very small intensity. Labelling is consistent with figure \ref{fig:schemes_D2_sigma}.}
    \label{fig:D2_tr_sigma_forbidden}
\end{figure}

\noindent With the slopes exhibited before, we can estimate that the frequency detuning between the two groups of r1-transitions is approximately 
\begin{equation}
\Delta\omega_{r1}^{D_2^{(\sigma)}} \approx \left| 2\mu_BB_z\right|\, .
\end{equation}
leading to $\Delta\omega_{r1}^{D_2^{(\sigma)}} \approx 25.19$ GHz, perfectly consistent with the value $25.15$ GHz on the top panel of figure \ref{fig:abs_D2_sigma_HPB}, the difference coming for the fact that hyperfine splittings are neglected. Similarly, we can estimate $\Delta\omega_{r2}^{D_2^{(\sigma)}}$ to be 
\begin{equation}
\Delta\omega_{r2}^{D_2^{(\sigma)}} \approx \left|\frac{10\mu_BB_z}{3}\right|
\end{equation}
that is to say approximately $41.98$ GHz, to compare with the value $42.07$ GHz of figure \ref{fig:abs_D2_sigma_HPB}. Reaching hyperfine Paschen-Back regime combined with sub-Doppler spectroscopic techniques allows to form narrow resonances far-detuned from the resonant frequency of the transitions. Choosing the right alkali for such studies is a key point since $B_0$ varies a lot. Hyperfine splittings and natural linewidths are also to take into account depending on the desired resolution. 
We will now focus on the so-called forbidden transitions, obeying $\Delta F = \pm 2$. Their intensities are plotted on figure \ref{fig:D2_tr_sigma_forbidden}.

\begin{figure}
    \centering
    \includegraphics[scale=0.55]{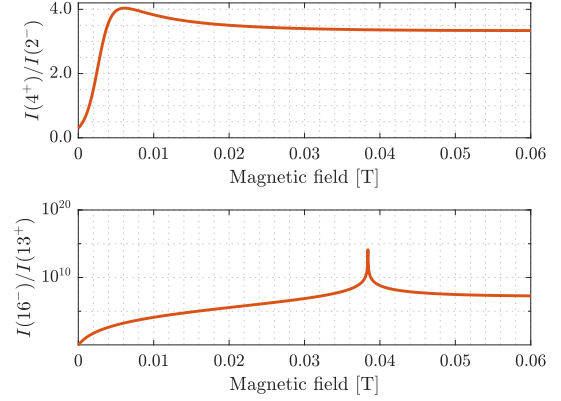}
    \caption{(color online) Intensity ratio between the two strongest MI-transitions of their group. Top panel: MI-transitions obeying $\Delta F = +2$ (group G2$^+$, $I(4^+)/I(2^-)$). Bottom panel: MI-transitions obeying $\Delta F = -2$ (group G2$^-$, $I(16^-)/I(13^+)$).}
    \label{fig:ratios}
\end{figure}

\noindent At first glance, what can be observed on figure \ref{fig:D2_tr_sigma_forbidden} is the overall difference in intensity between the two polarizations: while v-transition $2$ ($\sigma^-$, hereafter denoted $2^-$) peaks at approximately $0.16$, v-transition $4^+$ peaks at nearly 0.6 (both panels have the same y-scale). These transitions arise from the same ground state $\ket{1,-1}$, and $2^-$ (resp. $4^+$) goes to $\ket{3,-2}$ (resp. $\ket{3,0}$) due to the selection rule on $\Delta m_F$ and obey $\Delta F = +2$. The inverse behavior can be observed for v-transitions $16^-$ and $13^+$: their ground states are different, respectively $\ket{2,-1}$ and $\ket{2,+1}$, but they have the same excited state $\ket{0,0}$. They obey $\Delta F = -2$. Transition $16^-$ reaches a little more than $0.14$, while transition $13^+$ only reaches $5\times 10^{-3}$. To observe this behavior more clearly, we will plot the ratios $I(4^+)/I(2^-)$ and $I(16^-)/I(13^+)$. They are presented on figure \ref{fig:ratios}. We call group G2$^+$ the transitions obeying $\Delta F = +2$ and group  G2$^-$ the transitions obeying $\Delta F = -2$.

\begin{figure}
    \centering
    \includegraphics[scale=0.55]{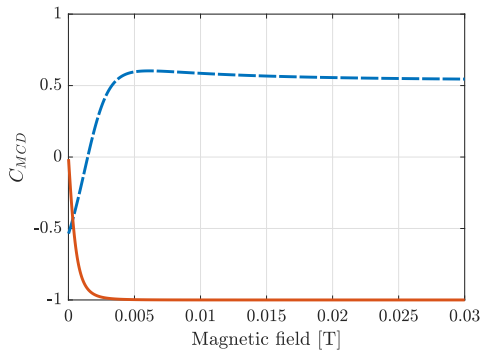}
    \caption{(color online) Figure of merit coefficient $C_\text{MCD}$ for each group of transitions. Blue (dashed) line: $(I(4^+)-I(2^-))/(I(4^+)+I(2^-))$. Orange (continuous) line: $(I(13^+)-I(16^-))/((13^+)+I(16^-))$. }
    \label{fig:mcdcoef}
\end{figure}

\noindent The ratio $I(4^+)/I(2^-)$ has a maximum of approximately $4$, occuring for $B_z \approx 6.12$ mT, whereas on the bottom panel  $I(16^-)/I(13^+)$ reaches as much as $1.25\times 10^{14}$ for $B_z \approx 38.41$ mT. This phenomenon is referred to as MI-Circular Dichroism (MCD) or frequency-controllable Circular Dichroism: for $\Delta F = +2$ the strongest $\sigma^+$ transition intensity is more than $4$ times higher than the strongest $\sigma^-$ one and for $\Delta F = -2$, $\sigma^-$ transitions are stronger than $\sigma^+$. By varying the laser frequency and its polarization, it is thus possible to choose whether $\sigma^+$ or $\sigma^-$ MI-transition will dominate. The figure of merit coefficient allows, for each group, to better visualize the dichroism phenomenon. It is defined as
\begin{equation}
C_\text{MCD} = \frac{I^+ - I^-}{I^+ + I^-}\label{eq:mcd}
\end{equation}
and is represented for the aforementioned transitions on figure \ref{fig:mcdcoef}. 

\noindent The coefficient defined in relation \eqref{eq:mcd} can be interpreted as follows: $C_\text{MCD} = 0$ represents equality between the intensities of the two strongest MI-transitions (one per polarization) of each group. $C_\text{MCD} > 0$ (resp. $C_\text{MCD} < 0$) means that the strongest transition occurs for $\sigma^+$ (resp. $\sigma^-$) thus leading to an asymmetry. $C_\text{MCD} = -1$, which is the case for the orange curve of figure \ref{fig:mcdcoef} (transitions belonging to group G2$^-$), means complete vanishing of the $\sigma^+$ transition and is reached for $B_z > 3$ mT. This type of MCD is referred to as Type 1 MCD and has been thoroughly described and observed for other alkali (Rb and Cs, mainly) in \cite{klingerthese} and \cite{tonoyanEPL}.

Moreover, it was recently demonstrated in \cite{sargsyanPLA} (theoretically and experimentally, by using a Cs-filled nanocell) that even among the strongest MI-transitions arising from different ground states for each circular polarization, the strongest MI-transition obeying $\Delta F = +2$ is always bigger than the strongest MI-transitions obeying $\Delta F = -2$. This is verified here: transition $4^+$ ($\ket{1,-1}\rightarrow \ket{3,0}$) is stronger than $16^-$  ($\ket{2,1} \rightarrow \ket{0,0}$). Numerically, here is it verified for the range $2.4$ mT - $1$ T. As mentioned in \cite{sargsyanPLA}, calculations suggest that Type 1 and Type 2 MCD could be observed for other first fundamental series of $nS \rightarrow nP$ $D_2$ lines, where MI-transitions also occur.

\section{Conclusion}
In this paper, we have performed a complete theoretical description of the behavior of a sodium vapor confined in a nanometric-thin cell for a wide range of magnetic field (varying up to $1$ Tesla) and for three incident laser polarizations (linear, left and right-circular). While the Zeeman structure of sodium is identical to the one of $^{87}\text{Rb}$ or ${}^{39}\text{K}$, several changes can be reported such as $B_0$ and the hyperfine splittings (leading to slight changes in the behavior of transition intensities and transition frequencies). As mentioned before, the natural linewidth $\Gamma_\text{nat}$ of sodium is twice bigger than for other alkalis and leads to much broader absorption lines. However, for the same temperature the vapor pressure of sodium is much smaller than heavier alkali atoms ($\approx 10^{-7}$ torr, whereas for cesium it is $\approx6\times10^{-4}$ torr and $\approx 2.5\times10^{-4}$ torr for ${}^{85}\text{Rb}$). This leads to smaller collisional broadening, but also makes the transmitted and reflected signal smaller thus either a bigger temperature or more sensitive detectors would be required to record the signal when performing experiments. 
As an addition, we provided in this paper rough estimates of the detuning (with respect to the resonant frequency of the hyperfine transitions $F_g = 1 \rightarrow F_e = 1$ (for $D_1$ line) and $F_g = 1 \rightarrow F_e = 0$ (for $D_2$ line)) between the various groups of transitions that remain present in the spectra when the hyperfine Paschen-Back regime is reached. We highlighted for the first time the appearance of Type 1 and Type 2 Circular Dichroism in sodium vapors and provided high-resolution absorption spectra of sodium Zeeman transitions depending on the value of the external magnetic field. Complete description and understanding of all these magneto-optical processes are of utmost importance for further applications, for example in optical magnetometry. Upcoming experiments involving nanocells are planned at the Institute for Physical Research to provide an experimental verification of these results. Complete agreement between experiments and theory is expected as it was proven for all other alkalis (except lithium, for which it is very hard to fabricate a nano-cell).

\newpage
 \bibliographystyle{elsarticle-num} 
 \bibliography{biblio}

\end{document}